\documentclass[12pt]{article}
\usepackage{amsmath,amssymb,amsthm,color}
\usepackage{graphicx}
\usepackage{cite}
\usepackage{epsfig}
\renewcommand{\baselinestretch}{2}

\begin{document}

%\begin{frontmatter}
%
\title{Configuration-enriched magnetoelectronic spectra of AAB-stacked trilayer graphene \\}
\author{
\small Thi-Nga Do $^{a}$, Chiun-Yan Lin$^{a,*}$,Yi-Ping Lin $^{a,*}$, Po-Hsin Shih $^{a,*}$, Ming-Fa Lin$^{a,*}$ $$\\
\small  $^a$Department of Physics, National Cheng Kung University, Tainan, Taiwan 701\\
 }
\renewcommand{\baselinestretch}{1}
\maketitle

\renewcommand{\baselinestretch}{1.4}
\begin{abstract}

We developed the generalized tight-binding model to study the magneto-electronic properties of AAB-stacked trilayer graphene. Three groups of Landau levels (LLs) are characterized by the dominating subenvelope function on distinct sublattices. Each LL group could be further divided into two sub-groups in which the wavefunctions are, respectively, localized at 2/6 (5/6) and 4/6 (1/6) of the total length of the enlarged unit cell. The unoccupied conduction and the occupied valence LLs in each sub-group behave similarly. For the first group, there exist certain important differences between the two sub-groups, including the LL energy spacings, quantum numbers, spatial distributions of the LL wavefunctions, and the field-dependent energy spectra. The LL crossings and anticrossings occur frequently in each sub-group during the variation of field strengths, which thus leads to the very complex energy spectra and the seriously distorted wavefunctions. Also, the density of states (DOS) exhibits rich symmetric peak structures. The predicted results could be directly examined by experimental measurements. The magnetic quantization is quite different among the AAB-, AAA-, ABA-, and ABC-stacked configurations.

\vskip 1.0 truecm
\par\noindent

\vskip1.0 truecm

\par\noindent  * Corresponding author. 
{~ Tel:~ +886-6-275-7575.}\\~{{\it E-mail addresses}: l28981084@mail.ncku.edu.tw (C.Y. Lin), mflin@mail.ncku.edu.tw (M.F. Lin)}
\end{abstract}

\pagebreak
\renewcommand{\baselinestretch}{2}
\newpage

{\bf 1. Introduction}
\vskip 0.3 truecm

Graphene, being a one-atom-thick layer of carbon atoms densely packed in a two-dimensional honeycomb lattice, has attracted a lot of theoretical and experimental research \cite {MKoshino, ADMartino, JSari, YHHo, MMKruczynski, YPLin, YKHuang, SHRSena, AKumar, YBZhang, CLKane, KSNovoselov, KSNovoselov2, VPGusynin, FGuinea, XDu, JWJiang, MMShokrieh, JULee, QLu, CLee}. It exhibits many unusual physical properties, e.g., a rich magnetic quantization \cite {MKoshino, ADMartino, JSari, YHHo, MMKruczynski, YPLin, YKHuang, SHRSena}, half-integer Hall effect \cite {AKumar, YBZhang, CLKane, KSNovoselov, KSNovoselov2, VPGusynin, FGuinea, XDu}, high Young's modulus \cite {JWJiang, MMShokrieh, JULee, QLu, CLee}, high Fermi velocity ($10^6$ m/s), and others. Graphene could play an important role in technological applications such as electric circuits \cite {LHuang, YMLin}, field-effect transistors \cite {IMeric, FNXia}, light-emitting diodes \cite {JBWu, THHan, GHJo}, solar cells \cite{JBWu2, JDRoyMayhew, NLYang, WJHong}, and durable touch screens \cite{SKBae, JWang}. Due to the hexagonal symmetry with a rotational angle of 60$^\circ$, a non-doped  graphene is a zero-gap semiconductor with a vanishing density of state at the Fermi level. The essential electronic properties can be drastically changed by the layer number \cite{HHibino, EHMFerreira, AHCNeto}, stacking configuration \cite{AHCNeto, CYLin, JSLee, MAoki, FGuinea2, KFMak}, magnetic field \cite{YZhang, MOGoerbig}, electric field \cite{CLLu, EVCastro, KFMak2}, dopping \cite{DMBasko, CCasiraghi}, mechanical strain \cite{JELee, SMChoi, JHWong}, and temperature variation \cite{YWTan, VVCheianov}. Few- and multi-layer graphenes have been successfully produced by experimental methods such as exfoliation of highly orientated pyrolytic graphite \cite{MJWebb, ZLiu, ZYRong, JMCampanera}, metalorganic chemical vapour deposition (MOCVD) \cite{CTEllis, JHHwang, ZYJuang, DRLenski, BJayasena, CHLui}, chemical and electrochemical reduction of graphene oxide \cite{KFMak3, LYZhang, LBBiedermann}, and arc discharge \cite{YPWu, ZSWu}. There exist important stacking configurations, including AAB \cite{ZYRong, JMCampanera, LBBiedermann}, ABC \cite{KSNovoselov3, BLalmi, CHLui, KFMak3, LYZhang, LBBiedermann}, AAA \cite{JHHwang, ZYJuang}, ABA \cite{BLalmi, CTEllis, CHLui, LBBiedermann}, and twisted \cite{JHHwang, LBBiedermann} and  turbostratic ones \cite{DRLenski}. The interlayer atomic interactions and stacking configurations induce the rich electronic properties of graphene. In this work, we investigate the complex relationship between the magnetic quantization and the interlayer atomic interactions in AAB-stacked trilayer graphene by using the generalized tight-binding model.

The low-lying energy dispersions of monolayer graphene possess a pair of linear bands intersecting at $E_F$. These energy bands are further quantized by the application a uniform perpendicular magnetic field $B=B_0\hat{z}$ \cite{SWu}. The magneto-electronic spectrum of the isotropic Dirac cones satisfies a simple $E^{c,v}\propto \pm\sqrt{nB_0}$ relationship, where $n$ is the quantum number, and $c$ and $v$ denote the conduction and valence Landau levels (LLs), respectively. The band structure and the LLs of monolayer graphene have been verified by a number of experimental methods \cite{VWBrar, ADeshpande, KSNovoselov3}. Bilayer graphene, being held together by Van der Waals interactions, can exhibit the highly symmetric AA and AB configurations. The former and the latter have, respectively, two pairs of linear and parabolic bands. The two isotropic Dirac-cones in the AA-stacked system are magnetically quantized into two groups of LLs with monolayer-like wavefunctions and energy spectrum \cite{YHHo2}. Also, the AB-stacked system presents two groups of LLs, with each LL having a single-mode wavefunction in the absence of LL anticrossing \cite{YHLai}. The field-dependent LL energy spectrum of AB-stacked bilayer graphene is different from that of monolayer graphene. It includes a few intergroup LL anticrossings at large field strengths ($B_0>$ 100T). The band structures \cite{VWBrar, KSNovoselov4, TOhta} and the first LL groups \cite{KSNovoselov4} have been experimentally verified for bilayer graphene.

The magneto-electronic properties of AAA- and ABA- stacked trilayer graphenes can be regarded as combinations of those of monolayer graphene \cite{CYLin}, and AB-stacked bilayer and monolayer graphenes \cite{CYLin}, respectively. Distinctly, the zero-field band structure of ABC-stacked trilayer graphene exhibits three pairs of energy dispersions: linear cones, sombrero-shaped bands, and parabolic bands \cite{YPLin, CYLin}. Accordingly, the $B_0$-dependent energy spectrum has both intergroup and intragroup LL anticrossings, in which the latter are arised from the sombrero-shaped energy bands. The band structures and the first LL groups of these stacking systems have been examined by numerous experimental measurements \cite{LYZhang, CHLui2, FZhang, TTaychatanapat}. As for the stacking symmetry, the AAB-stacked system is in sharp contrast with the AAA- and  ABA-stacked ones, but is also different from the BAC-stacked one. Clearly, the AAB- and ABC-stacked trilayer graphenes can not be regarded as a superposition of two or three sub-systems.

We develop the generalized tight-binding model, based on the subenvelope functions on the distinct sublattices, to study the rich electronic properties of AAB-stacked trilayer graphene. This work shows that the AAB-stacked configuration presents an abnormal band structure, in which there are pairs of oscillatory, sombrero-shaped and parabolic bands, which are different from those of the other stacking systems mentioned above. The special band structure could be verified by angle-resolved photoemission spectroscopy (ARPES) \cite{MSprinkle}. The LL evolution under a magnetic field reveals a complex pattern of LL anticrossings and splittings, as a result of the specific interlayer atomic interactions derived from the full tight-binding model. The magnetoelectronic spectra are directly reflected in DOS which can be examined by experimental measurements using scanning tunneling spectroscopy (STS). The $B_0$-dependent energy spectra exhibit intragroup and intergroup LL anticrossings. Especially, the low-lying LLs anticross frequently, leading to the existences of non-well-behaved and perturbed LLs. The spatially dramatic changes of the LL wavefunctions in the anticrossings can be verified by scanning tunneling microscope (STM) measurements \cite{ZYRong, JMCampanera}, as done for 2D electron gas \cite{MFCrommie} and topological insulators \cite{ZHYang}. In this paper, the important differences among trilayer graphenes such as their energy band structures, LL splittings, LL ordering, and LL anticrossings, are investigated.

\newpage
\vskip 0.6 truecm
\par\noindent
{\bf 2. The generalized tight-binding model }
\vskip 0.3 truecm

The low-energy $\pi$-electronic structure of AAB-stacked trilayer graphene, mainly coming from the 2$p_z $ orbitals, is calculated with the generalized tight-binding model. The two sublattices in the $l$th $(l=1,2,3)$ layer are denoted as $A^{l}$ and $B^{l}$. The first two layers, shown in Fig. 1, are arranged in the AA-stacking configuration; that is, all carbon atoms have the same (x,y) projections. The third layer can be obtained by shifting the first (or the second) layer by a distance of $b$ along the armchair direction. In this system, the A atoms (black) have the same (x,y) coordinates, while the B atoms (red) on the third layer are projected at the hexagonal centers of the other two layers. The interlayer distance and the C-C bond length are, respectively, $d=3.37 \dot{A}$ and $b=1.42 \dot{A}$. There are six carbon atoms in a primitive unit cell. The low-energy electronic properties are characterized by the carbon $2p_{z}$ orbitals. The zero-field Hamiltonian, which is built from the six tight-binding functions of the $2p_{z}$ orbitals, is dominated by the intralayer and the interlayer atomic interactions $\gamma_i^{'}s$. There exist 10 kinds of atom-atom interactions corresponding to the 10 atomic hopping integrals which appear in the Hamiltonian matrix. $\gamma_{0}=-2.569  $ eV represents the nearest-neighbor intralayer atomic interaction; $ \gamma_{1}=-0.263$ eV, $\gamma_{2}=0.32 $ eV present the interlayer atomic interactions between the first and second layer; $ \gamma_{3}=-0.413$ eV, $\gamma_{4}=-0.177 $ eV, $ \gamma_{5}=-0.319$  eV are associated with the interlayer atomic interactions between the second and third layer; $ \gamma_{6}=-0.013  $ eV, $ \gamma_{7}=-0.0177 $ eV, and $ \gamma_{8}=-0.0319  $ eV relate to the interlayer atomic interactions between the first and third layer; and $ \gamma_{9}=-0.012 $ eV accounts for the difference in the chemical environment of A and B atoms. The hopping integrals $\gamma_1$, $\gamma_3$, and $\gamma_5$ belong to the vertical interlayer atomic interactions, while the others are non-vertical ones.

When applying a uniform perpendicular magnetic field, the unit cell becomes enlarged as indicated in Fig. 1. There appears an extra Peierls phase $G_{R}$ in the tight-binding functions. $G_{R}$ has the form of $ \frac{2\pi}{\phi_{0}}\int_R^r \vec{A}. d\vec{l} $, where $\vec{A}$ is the vector potential, and $\phi_{0}=hc/e$ is the flux quantum. The vector potential in the Landau gauge is chosen as $\vec{A}=(0,Bx,0)$. The Peierls phase has a period of $2\phi_{0}/ \phi=2R_{B}$ along the x-axis. Under the effect of the magnetic field, the unit cell becomes enlarged as a rectangle with $12R_{B}$ atoms included. There are $12R_B$ tight-binding functions, which are arranged in the sequence $\{|A_{1}^{1}\, |B_{1}^{1}\rangle, |A_{1}^{2}\rangle, |B_{1}^{2}\rangle, |A_{1}^{3}\rangle, |B_{1}^{3}\rangle, ..., |A_{2R_{B}}^{1}\, |B_{2R_{B}}^{1}\rangle, |A_{2R_{B}}^{2}\rangle, |B_{2R_{B}}^{2}\rangle, |A_{2R_{B}}^{3}\rangle, |B_{2R_{B}}^{3}\rangle\}$. The superscript Hamiltonian is a $12R_{B} \times 12R_{B} $ matrix, in which the non-zero matrix elements can be presented by the equations below.
\begin {equation*}
\langle B_{j}^{1} |H|A_{i}^{1}\rangle = \gamma_{0} \sum  { \frac{1}{N} exp [ i\vec{k}.(\vec{R}_{A_{i}^{'}}- \vec{R}_{{B}_{j}^{'}}   )] exp[\frac{2i\pi}{\phi_0} (G_{  \vec{R}_{{B}_{j}^{'}}} -  G_{  \vec{R}_{{B}_{j}^{'}}})] }\\
\end{equation*}
\begin{equation}
=\gamma_{0}t_{1,i}\sigma_{i,j} + \gamma_{0}q\sigma_{i,j+1} 
\end {equation}
 \begin{equation}
\langle B_{j}^{2} |H|A_{i}^{2}\rangle 
=\gamma_{0}t_{1,i}\sigma_{i,j} + \gamma_{0}q\sigma_{i,j+1} 
\end {equation}

 \begin{equation}
\langle B_{j}^{3} |H|A_{i}^{3}\rangle 
=\gamma_{0}t_{3,i}\sigma_{i,j-1} + \gamma_{0}q\sigma_{i,j} 
\end {equation}

 \begin{equation}
\langle A_{j}^{2} |H|A_{i}^{2}\rangle =\langle A_{j}^{3} |H|A_{i}^{3}\rangle
=\gamma_{9}\sigma_{i,j}
\end {equation}

 \begin{equation}
\langle A_{j}^{2} |H|A_{i}^{1}\rangle =\langle B_{j}^{2} |H|B_{i}^{1}\rangle =\gamma_{1}\sigma_{i,j}
\end {equation}

 \begin{equation}
\langle B_{j}^{1} |H|A_{i}^{2}\rangle =\langle A_{j}^{1} |H|B_{i}^{2}\rangle =\gamma_{2} t_{1,i}\sigma_{i,j}
\end {equation}

 \begin{equation}
\langle A_{j}^{3} |H|A_{i}^{2}\rangle =\gamma_{3}\sigma_{i,j}
\end {equation}

 \begin{equation}
\langle B_{j}^{3} |H|B_{i}^{2}\rangle =\gamma_{5}t_{2,i}\sigma_{i,j}
\end {equation}

 \begin{equation}
\langle B_{j}^{2} |H|A_{i}^{3}\rangle 
=\gamma_{4}t_{1,i}\sigma_{i,j} + \gamma_{4}q\sigma_{i,j+1} 
\end {equation}

 \begin{equation}
\langle A_{j}^{2} |H|B_{i}^{3}\rangle 
=\gamma_{4}t_{3,i}\sigma_{i,j-1} + \gamma_{4}q\sigma_{i,j} 
\end {equation}

 \begin{equation}
\langle A_{j}^{3} |H|A_{i}^{1}\rangle 
=\gamma_{6}\sigma_{i,j} 
\end {equation}

 \begin{equation}
\langle B_{j}^{3} |H|B_{i}^{1}\rangle 
=\gamma_{8}t_{2,i}\sigma_{i,j} + \gamma_{8}q\sigma_{i,j+1} 
\end {equation}

 \begin{equation}
\langle B_{j}^{1} |H|A_{i}^{3}\rangle 
=\gamma_{7}t_{1,i}\sigma_{i,j} + \gamma_{7}q\sigma_{i,j+1} 
\end {equation}

 \begin{equation}
\langle A_{j}^{1} |H|B_{i}^{3}\rangle 
=\gamma_{7}t_{3,i}\sigma_{i,j-1} + \gamma_{7}q\sigma_{i,j} 
\end {equation}
The four independent phase terms are:\\
$t_{1,j} = exp \{ i[ -(k_{x}b/2) - (\sqrt {3}k_{y}b/2) +\pi\phi (j-1+ 1/6) ] \}    \\
+ exp \{ i[ -(k_{x}b/2) + (\sqrt {3}k_{y}b/2) -\pi\phi (j-1+ 1/6) ] \} $  \\
$t_{2,j} = exp \{ i[ -(k_{x}b/2) - (\sqrt {3}k_{y}b/2) +\pi\phi (j-1+ 3/6) ] \}    \\
+ exp \{ i[ -(k_{x}b/2) + (\sqrt {3}k_{y}b/2) -\pi\phi (j-1+ 3/6) ] \} $  \\
$t_{1,j} = exp \{ i[ -(k_{x}b/2) - (\sqrt {3}k_{y}b/2) +\pi\phi (j-1+ 5/6) ] \}    \\
+ exp \{ i[ -(k_{x}b/2) + (\sqrt {3}k_{y}b/2) -\pi\phi (j-1+ 5/6) ] \} $  \\
$q=exp\{ ik_{x}b\}$.\\
In order to deal with the small values of magnetic field strength and a huge $R_B$, we arrange the Hamiltonian in a band-like symmetric matrix. The Landau wave functions, which are investigated to identify the spatial distributions of the Landau levels, can be expressed as 
\begin{equation}
|\psi\rangle = \sum_{l=1}^{3}\sum_{m=1}^{2R_{B}} (A_m^l | A_m^l \rangle    + B_m^l | B_m^l \rangle    ),
\end {equation}
where $A_m^l (B_m^l)$ are the subenvelope functions presenting the amplitude of the tight-binding functions based on the $m$th A (B) atom at the $l$th layer in the unit cell.

\vskip 0.6 truecm
\par\noindent
{\bf 3. The zero-field band structure and quantized Landau levels}
\vskip 0.3 truecm

The zero-field band structure of AAB-stacked trilayer graphene consists of three pairs of conduction and valence subbands labeled $S_1^{c,v}$, $S_2^{c,v}$, and $S_3^{c,v}$, as shown by the solid curves in Fig. 2(a). Near the Fermi energy, the two subbands which belong to the first pair, $S_1^{c,v}$, have strong oscillatory energy dispersions. The conduction subband starts to increase from a local minimum value of about 4 meV at the K point (the corner of the first Brillouin zone), along the KM and K$\Gamma$ directions. After reaching a local maximum value of about 58 meV, it decreases until reaching a local minimum energy again (4 meV), and then grows steadily. The curvature of the valence subband is in the opposite direction, which is almost symmetric to the conduction subband about $E_F$. The first pair of subbands, with three constant energy contours within $\pm$ 58 meV and a narrow gap $ E_g\sim 8$ meV in between them (Fig. 2(b)), is special in that it has never appeared in any other stacking configuration. The triple-degenerate states are not suitable for low-energy expansion, indicating that the effective-mass model can not further deal with the magnetic quantization of these energy bands. The second pair of subbands, $S_2^{c,v}$, has a sombrero-shaped and a local energy minimum (maximum) and maximum (minimum), situated at around 0.24 $(-0.24)$ eV and 0.26 $(-0.26)$ eV, respectively. The energy difference between the two extreme points is quite narrow, being only about 20 meV. Located away from $E_F$, the third pair of subbands, $S_3^{c,v}$, consists of monotonic parabolic bands with a minimum (maximum) value of about 0.49 $(-0.49)$ eV. The above-mentioned features of the low-lying energy bands are consistent with those by the first-principle calculations (dashed curves), clearly indicating that the complex interlayer amomic interactions used in the generalized tight-binding model are suitable.

As for the low-lying band structure, there are important differences among trilayer graphenes with distinct stacking configurations. The AAA-, ABA-, and ABC-stacked trilayer graphenes possess special band structures with three pairs of linear bands, a pair of linear bands and two pairs of monotonic parabolic bands, and a pair of sombrero-shaped bands and two pairs of parabolic subbands, respectively. The band structures of the AAA and ABA stackings can be, respectively,  regarded as the superposition of monolayer graphene, and monolayer and bilayer graphenes. In contrast, ABC- and AAB-stacking configurations exhibit special energy band structures, which are not similar to those of monolayer and bilayer systems. It should be noted that the energy band structure of AAB-stacking system exhibits an extremely small gap, as indicated in Fig. 2(b). That is to say, while the three stacking trilayer graphenes are gapless 2D semimetals with a slight overlap between the valence and conduction bands, the AAB-stacked system is a narrow-gap semiconductor. 

A perpendicular magnetic field can quantize electronic states into dispersionless LLs with high degeneracy. AAB-stacked trilayer graphene presents rich magnetoelectronic properties. For the $(k_x=0, k_y=0)$ state, each LL is two-fold degenerate in the absence of spin degeneracy; that is, the wavefunctions localized near 2/6 and 5/6 (4/6 and 1/6) are identical. The total carrier density, in which each LL can be occupied, is $D$=4$eB_0/hc$ per unit area. There exist three groups of LLs, each group can furthermore be divided into two sub-groups corresponding to the two localizations 2/6 and 4/6. Each group consists of unoccupied conduction and occupied valence LLs. The former and the latter are almost symmetric about $E_F$ and present similar behavior. We first discuss the LLs localized near 2/6 at $B_0=40$ T, being characterized by the subenvelope functions on the six sublattices shown in Figs. 3-5. The first, second and third groups are, respectively, initiated at 0, 0.24 and 0.55 eV for the conduction states, and at 0, -0.23 and -0.58 eV for the valence states. Near $E_F$, the first group starts to show up with the first three LLs (blue lines), which are very close to each other. The quantum numbers are determined by the $A^1$ sublattice with the dominating subenvelop function (Fig. 3). Starting from the middle LL, $n_1^{c,v}=0$, which is located almost right at $E_F$, the ordering then increases for the higher conduction states, and the lower valence states. In particular, the quantum number of the next conduction (valence) LL, which is placed at about $E^c=$10 meV ($E^v=-$5.2 meV), is assigned $n_1^c = 1$ ($n_1^v =1$), and so on. For the first group, the quantum number ordering is similar to that of monolayer graphene, while the LL spacing is irregular. The largest spacing between LLs is that of $n_1^{c,v}=1$ and $n_1^{c,v}=2$, which is about $ \Delta E^{c,v}=$ 40 meV; it then decreases with increasing quantum numbers. Apparently, there is no simple relationship between the LL energy and quantum number. This is in contrast to the relationships found in monolayer graphene and 2D electron gas, which can be described by $E^{c,v} \propto \sqrt {n^{c,v}}$ and $E(n) \propto n^{c,v}$, respectively.

The destruction of the inversion symmetry in AAB-stacked trilayer graphene results in certain important differences for the LL wavefunctions near 4/6 and 2/6, including differences in the LL spacing, quantum number and spatial distribution of the wavefunctions. At $B_0=40$ T, the first, second and third LL groups near the 4/6 center are, initiated at 0, (0.26 eV, $-$0.24 eV) and $\pm$0.52 eV, as indicated in Fig. 4(a)-4(c), respectively. For the low-lying LLs, it is not easy to define the quantum numbers due to the non-well-behaved spatial distributions of the LL wavefunctions. Particularly, the subenvelope functions oscillate abnormally as a result of the complex LL anticrossings (details in Fig. 6(c)). The eleven LLs nearest to $E_F$, except for the $n=0$ one, are arranged in pairs with very small spacings, in which the LL crossings or anticrossings are clearly revealed (Fig. 4(a)). At low-lying energy levels, the LL wavefunctions are well-behaved only when the applied field strength is sufficiently large (the available magnetic field in experiment is nowadays up to 80 T \cite{BFauque}), as shown in Fig. 5(a)-5(c) for $B_0=80$ T. The dominating subenvelope functions of the sublattice $B^3$ are used to define the quantum numbers. Similarly to those near 2/6 center, the conduction and valence LLs are also almost symmetric about the $n=0$ LL, which is located right at $E_F$. However, the quantum number ordering is slightly different to that of monolayer graphene. Particularly, the next unoccupied (occupied) LLs are assigned, respectively, $n_1^c=2$ ($n_1^v=2$), $n_1^c=3$ $n_1^v=3$, $n_1^c=1$ ($n_1^v=1$), and so on.

Concerning the second group, all the LL wavefunctions are single-mode at $B_0=40$ T. Also, the LL quantum numbers are determined by the dominating $ B^1 $ sublattice for the wavefuntions distributed around 2/6 and 4/6 (green color in Figs. 3(b) and 4(b)). The assigned quantum numbers of the former are $n_2^{c,v}=1$, $n_2^{c,v}=0$, $n_2^{c,v}=2$, $n_2^{c,v}=3$, ... in the order of increasing (decreasing) energy for the conduction (valence) LLs, similarly to those of the latter ($n_2^{c,v}=1$, $n_2^{c,v}=0$, $n_2^{c,v}=2$, $n_2^{c,v}=3$). It should be noted that when the field strength is sufficiently large, the quantum number ordering becomes equivalent to that of monolayer graphene, as shown in Fig. 5 (b) near the 4/6 center for $B_0=80$ T (the available magnetic field in experiment is nowadays up to 80 T \cite{BFauque}). On the other hand, the third group exhibits a normal quantum number sequence for two localization centers, similarly to that of monolayer graphene. That is to say, the ordering is $n_3^{c,v}=0$, $n_3^{c,v}=1$, $n_3^{c,v}=2$, $n_3^{c,v}=3$, and so on. The LL wavefuntions based on the dominating subenvelope function $A^2$, are almost identical to those of monolayer graphene, and independent of the field strength.

The $B_0$-dependent energy spectra are very useful in understanding the rich magnetic quantization in AAB-trilayer graphene. The LL energies exhibit monotonic and entangled spectra, in which the multi-crossings and -anticrossings occur frequently (Fig. 6). For all the LLs of the first group localized at 2/6, the quantum numbers have a normal sequence when the field strength is very large ($B_0> 40$ T). Moreover, the conduction (valence) LL energies monotonously decline (grow) when the field strength is reduced to $B_0=40$ T. With a further decrease of field strength, the $B_0$-dependent energy spectrum becomes oscillatory, which leads to the frequent and pronounced multi-crossings and -anticrossings. If two different multi-mode LLs simultaneously possess the same mode, they are forbidden to cross each other. For example, the three conduction LLs, $n_1^{c}=1$, $n_1^{c}=4$ and $n_1^{c}=7$ continuously anticross, as indicated in Fig. 6(b). In particular, the anticrossing between the former two occurs in the range of $B_0\sim\,19-21$ T, while in the latter two they occur between $B_0\sim\,9-11$ T; the anticrossing between the first and the third conduction LLs appears at $B_0\sim\,11-13$ T. The LL anticrossings clearly indicate the unusual sequence of quantum numbers. For $B_0 >$ 21 T, the main $n_1^c=1$ mode has a normal distribution on the dominating sublattice, in which the side modes of $n_1^c=4$ and $n_1^c=7$ are faint. With the decreasing of $B_0$, the $n_1^c=1$ main mode declines and the $n_1^c=4$ side mode quickly reaches the maximum at $B_0 = 20$ T in the range of $E^c\sim\, 20-40$ meV. At the center of the anticrossing regions, their subenvelop functions have the same $n=1$ and $n=4$ modes, which forbids them to have the same energy. Similarly, the $n_1^c=4$ and $n_1^c=7$ LLs anticross in the range of $E^c\sim\, 15-35$ meV. The anticrossing center is at  $B_0\sim\,10$ T where the comparable $n=4$ and $n=7$ modes exist in two subenvelope functions. Moreover, the other conduction LLs exhibit similar $B_0$-dependent energy spectra. Particularly, there are other pairs of LLs, such as $n_1^c=2$ and $n_1^c=5$, $n_1^c=3$ and $n_1^c=6$, ... , for which the LL anticrossings appear, respectively, in the ranges of 16 T $<B_0<$ 18 T ($E^c\sim\,18-38$ meV), 13 T $<B_0<$ 15 T ($E^c\sim\,17-37$ meV), and so on. The LL anticrossings form a wide stateless region which is indicated in Fig. 6(b). Generally, the main modes of anticrossing LLs are different by 3 at large field strengths and 6 at small ones.

Furthermore, there exists another region of LL anticrossings which is indicated in Fig. 7(a). These anticrossings happen at $B_0<$ 19 T and form very narrow in-between gaps. For example, the $n_1^c=1$ and $n_1^c=7$ LLs are forbidden to cross in the energy range of  $E^c\sim\, 34-36$ meV (Fig. 7(a)). Likewise, the other pairs of LLs, such as ($n_1^c=2$ and $n_1^c=8$  ) ($E^c\sim\, 39-40$ meV), ($n_1^c=3$ and $n_1^c=9$  ) ($E^c\sim\, 41-41.5$ meV), anticross at about $B_0=$ 11 T and $B_0=$ 10.6 T, respectively. With a further decrease of $B_0$, the $n_1^c=4$ and $n_1^c=10$ ($n_1^c=5$ and $n_1^c=11$) LLs anticross each other at around $B_0=$ 6 T ($B_0=$ 5 T) in the range of $E^c\sim\, 36-38$ meV ($E^c\sim\, 40-42$ meV), and so on. In short, for each pair of anticrossing LLs, the main modes differ by 6. In addition to the intragroup anticrossings of the LLs in the first group, the second group LLs also avoid crossing each other at smaller field strengths ($B_0 < 5$ T) due to the sombrero-shaped energy dispersions \cite{YPLin, CYLin} shown in Fig. 7(b). The third group consists of monotonic LLs without intragroup anticrossings.

The intergroup LL anticrossings occur among all three groups at sufficiently large field strengths. For instance, the $n_1^{v}= 3$ LL anticrosses the $n_2^{v}=0$ and $n_2^{v}=3$ ones continuously at $B_0\sim\,75-97$ T, in the range of, respectively, $E^v\sim\,$-23$-$18 meV and $E^{v}\sim\,$-27$-$-23 meV, as shown in Fig. 7(c). In general, the main modes of anticrossing LLs are different by 3m, with m being an integer. It should be noted that the interlayer atomic interaction $\gamma 5$ between the non-vertical sites in the second and third layers induces both intergroup and intragroup LL anticrossings. This is similar to ABC-stacked trilayer graphene \cite{CYLin}.

Significantly, near $E_F$, the $n=0$ LL presents the unusual field-dependent LL wavefunctions, as indicated in Fig. 6(b). When the magnetic field is sufficiently large, e.g., $B_0>22$ T, both $n=0$ and $n^{c,v}=3$ LLs present the well-behaved wavefunctions. However, the $n=0$ LL is forbidden to cross the $n^{c}=3$ one  when the field strength is reduced to $B_0\sim 22$ T. Accordingly, there appear some $n=0$ mode wavefunctions near the turning point of the $n^c=3$ LL, which is located at $B_0\sim 20$ T and $E^c\sim 13$ meV. Besides, the $n^c=3$ and $n^v=3$ LLs also weakly anticross right at their extreme points, forming an exceedingly narrow gap in the anticrossing center. As the field strength is gradually decreased, the $n=0$ and $n^c=3$ LLs anticross each other again at $B_0\sim 16$ T. Therefore, three LLs in the range of $E^c\sim\,$-10$-15$ meV and $B_0\sim\,16-22$ T exhibit an entangled energy spectrum including the special anticrossings.

On the other hand, there exist certain critical differences between the field-dependent LL spectra at the 4/6 and 2/6 localization centers. For the first group, the intragroup LL anticrossings of the former appear in a wider magnetic-field range, 0 T $<B_0<80$ T (Fig. 6(c)). Specifically, the complex LL anticrossings of the $n=0$ and $n^{c,v}=3$ LLs are revealed in the range of larger field strengths ($B_0\sim\,32-45$ T) compared to that of the latter ($B_0\sim\,16-22$ T), as indicated in Figs. 6(b)-(c). The $n=0$ LL anticrosses the $n^v=3$ LL twice, leading to the fact that the zero-mode LLs only exist above $E_F$, which is in contrast to those near the 2/6 center. Moreover, the stateless gap in the anticrossing center of $n^c=3$ and $n^v=3$ LLs is about 5 meV, clearly larger than that at 2/6 localization.

The $B_0$-dependent LL wavefunctions of AAB-stacked trilayer graphene exhibit diverse real-space distributions. They can be verified by the spectroscopic-imaging STM measurement, as done for 2D electron gas \cite{MFCrommie} and topological insulators \cite{ZHYang}. From the measurements on the variations of the local density of states (DOS) in graphene planes, STM reveals a nodal structure corresponding to the well-behaved LL wavefunctions. For the hybridized wavefunctions due to the LL anticrossings in AAB-trilayer graphene, the results from the STM measurements are expected to be helpful in distinguishing the main mode and the side modes of the LLs. Moreover, the energy spectra and the internal structure of the wavefunctions could be directly examined by STM and STS measurements (discussed later), respectively, giving a useful identifiable picture of the LLs.

Reflecting the main characteristics of LL spectra, the DOS is defined as $D(\omega )=\underset{n^{c},n^{v}}{\sum} \int_{1stBZ}\delta (\omega - E^{c,v}(n,k))dk.$ The discrete LLs lead to many symmetric delta-function-like peaks in the DOS, where the peak intensities are proportional to the LL degeneracy \cite{DLMiller, GMRutter}. Such peaks correspond to the sharp structures in the differential conductance map of dI/dV$-$V from STS measurements. For the DOS of the AAB-stacked trilayer graphene at B=40 T, the first, second and third groups, respectively, have the onset conduction-states about 0, 0.26 and 0.52 eV, as shown in Fig. 8(a). Each group consists of two sub-groups, which are separated by a symmetry-breaking induced energy difference of about 10 meV. In the vicinity of the Fermi level, the peaks of the first groups are neither regularly sequenced nor do they follow a simple relationship as for monolayer graphene. This is due to the fact that the special quantization effects on the lowest subband leads to the multi-crossings and anticrossings within the band width. On the other hand, the second and third groups exhibit a normal sequence in the order of increasing energies. A crossover of LLs in different groups induces higher DOS, which is thus expected to exhibit stronger tunneling currents in experimental measurements.  

The DOS exhibits distinct characteristics among different stacking configurations, such as the peak sequence, intensity, energy and splitting. In AB-stacked trilayer graphene, peaks that follow the sequence in monolayer and bilayer graphenes are observed, as presented in Fig. 8(b). Similarly, for the AAB stacking, the half-intensity peaks near the Fermi level are attributed to the symmetry-broken structure, which leads to a considerable energy splitting of about 10 meV. However, this splitting is hardly observable at higher energies. The AA-stacked trilayer graphene exhibits three groups of monolayer-like sequence of peaks \cite{DLMiller, SJung, YJSong} located at energies described by the simple relationship $E^{c,v}\propto\sqrt{n^{c,v}B}$, as indicated in Fig. 9(a). On the contrary, the DOS of ABC-stacked trilayer graphene is distinct from that of monolayer and bilayer graphenes, as shown in Fig. 9(b). The exceptionally high peak at the Fermi level is a superposition of three peaks corresponding to the Dirac points; its intensity is proportional to the number of graphene layers. The essential differences of the DOS can be verifed by STS \cite{DLMiller, GMRutter, SJung, YJSong}; those profiles can then be used as a tool to identify the stacking configuration of graphene sheets.

\vskip 0.6 truecm
\par\noindent
{\bf 4. Conclusion }
\vskip 0.3 truecm

%\section{Conclusions}

The generalized tight-binding model is developed to investigate the rich magneto-electronic properties of AAB-stacked trilayer graphene. This system exhibits special energy bands and thus a rich magnetic quantization. The complex interlayer atomic interactions indicate that the low-energy expansion about the K point is not suitable in obtaining energy dispersions; therefore, the effective-mass model can not be used to achieve further magnetic quantization.
The three zero-field pairs of energy bands contain oscillatory, sombrero-shaped and parabolic dispersions, which could be examined by ARPES \cite{MSprinkle}.
They are magnetically quantized into three groups of LLs, defined by the dominating subenveloped functions. The field-dependent  energy spectra, in which the frequent intragroup and intergroup LL anticrossings happen simultaneously, are very complex.
There exist important differences between the LLs near 4/6 and 2/6 in terms of the LL splitting, quantum number, spatial distribution of wavefunctions, and anticrossing LL energy spectra.
Moreover, the electronic properties of AAB-stacked trilayer graphene sharply contrast to those of the AAA-, ABA- and ABC-stacked ones, such as state degeneracy, LL anticrossing behavior, initial energies of each group, quantum number ordering and wavefunction distributions.
This principally demonstrates the unique characteristics of the special AAB configuration.

The first and the second LL groups exhibit unusual energy spectra with continuous intragroup LL anticrossings.
Especially for the first group, each low-lying LL is forbidden to cross with the others more than once. 
The LL anticrossing region near $E_F$ has a wide stateless gap, while that of the other region is quite narrow.
The main modes of LLs during anticrossings are different by 3 and 6 for the former, and only 6 for the latter.
Furthermore, the intergroup LL anticrossings among three groups are revealed only at sufficiently large field strengths.
Remarkably, the DOS exhibits many symmetric delta-function-like peaks with an irregular sequence near $E_F$ due to the LL multi-anticrossings.
The web-like energy spectra and the seriously distorted wavefunctions could be examined by STS and STM experimental measurements, respectively.
The complex LL spectra are expected to induce the rich mechanical, excitonic, thermal and optical properties of the AAB-stacking system.

\par\noindent {\bf Acknowledgments}

This work was supported in part by the National Science Council of Taiwan,
the Republic of China, under Grant Nos. NSC 98-2112-M-006-013-MY4 and NSC 99-2112-M-165-001-MY3.
%\end{document}

\newpage
%{\large\bf References} \\
\renewcommand{\baselinestretch}{0.2}
%\begin{itemize}
%\newpage

%\end{document}
\newpage \centerline {\Large \textbf {FIGURE CAPTIONS}}

Fig. 1 - The interlayer atomic interactions and the geometric structure under a uniform magnetic field $B_0\hat{z}$. The shaded region corresponds to a rectangular unit cell. The first and second layers have the same (x, y) projections.

Fig. 2 - The (a) energy band structures of AAB-stacked trilayer graphene with (b) a narrow energy gap.

Fig. 3 - The distributions of the LL wavefunctions based on the dominating sublattices centered at the (a)-(c) 2/6 localization under $B_0=40$ T.

Fig. 4 - The distributions of the LL wavefunctions based on the dominating sublattices centered at the (a)-(c) 4/6 localization under $B_0=40$ T.

Fig. 5 - The distributions of the LL wavefunctions based on the dominating sublattices centered at the (a)-(c) 4/6 localization under $B_0=40$ T.

Fig. 6 - Three groups of field-dependent LL energy spectrum for (a) the 2/6 localization center; the low-lying LL spectra for (b) the 2/6 and (c) 4/6 centers.   

Fig. 7 - The (a) second region of intragroup LL anticrossings in the first group, the (b) frequent intragroup LL anticrossings of  the second group at low field strength, and the (c) intergroup LL anticrossings of the first and second groups.

Fig. 8 - The DOS of (a) AAB- and (b) ABA-stacked trilayer graphene at $B_0=40$ T.

Fig. 9 - The DOS of (a) AAA- and (b) ABC-stacked trilayer graphene at $B_0=40$ T.
\vskip0.5 truecm 
%\section{Figure Captures}

\begin{figure}[htb]
\centering\includegraphics[width=0.9\linewidth]{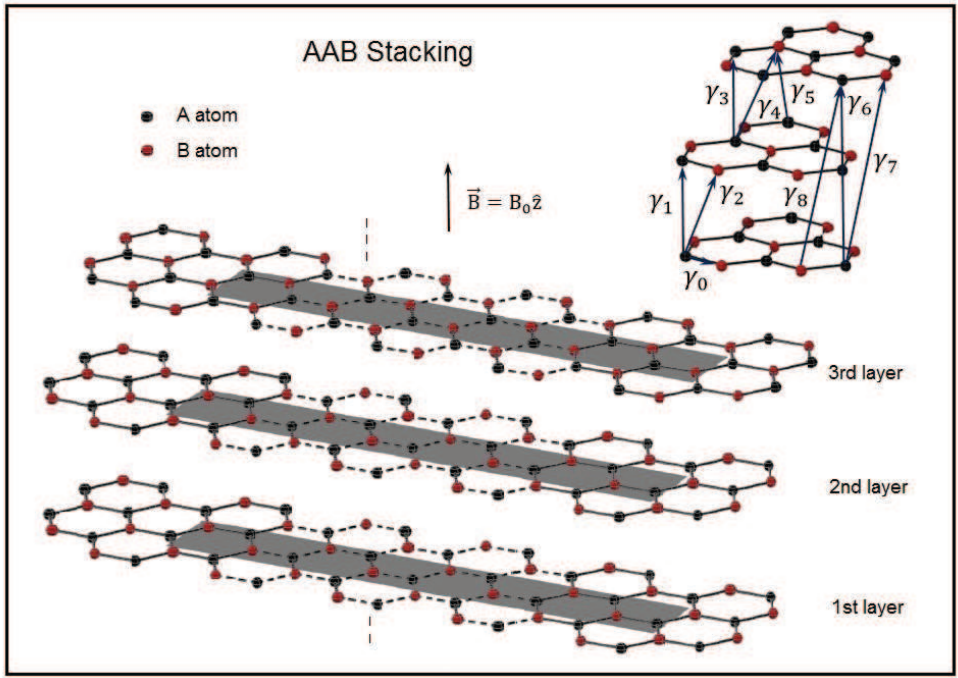}
\caption{The interlayer atomic interactions and the geometric structure under a uniform magnetic field $B_0\hat{z}$. The shaded region corresponds to a rectangular unit cell. The first and second layers have the same (x, y) projections.}
\end{figure}

\begin{figure}[htb]
\centering\includegraphics[width=0.6\linewidth]{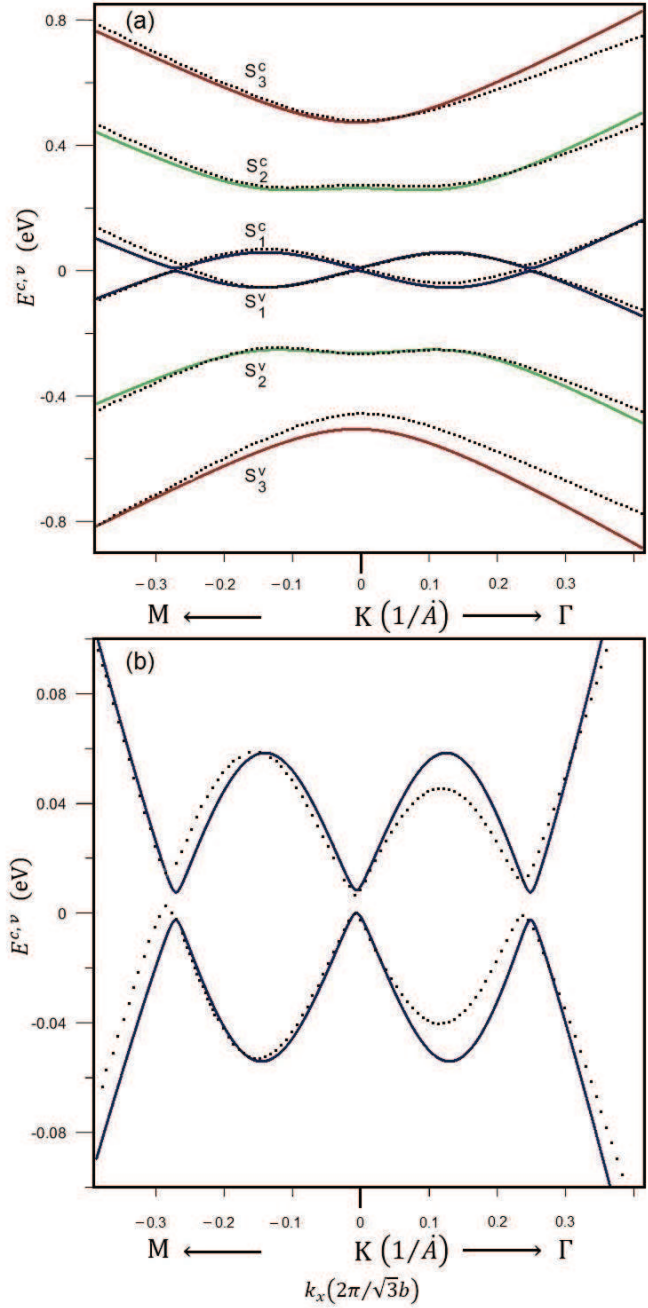}
\caption{The (a) energy band structures of AAB-stacked trilayer graphene with (b) a narrow energy gap.}
\end{figure}

\begin{figure}[htb]
\centering\includegraphics[width=0.9\linewidth]{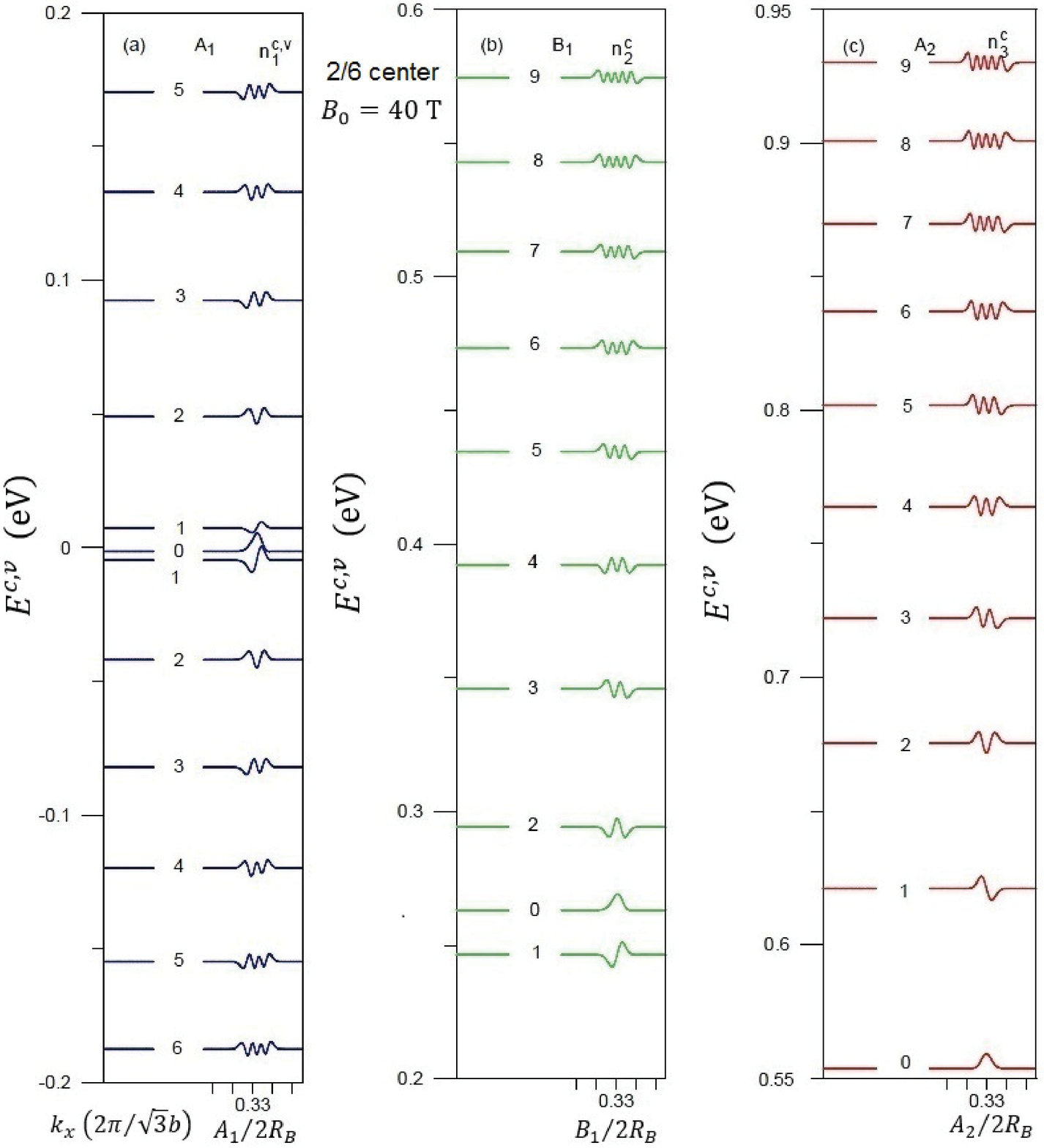}
\caption{The distributions of the LL wavefunctions based on the dominating sublattices centered at the (a)-(c) 2/6 localization under $B_0$ = 40 T.}
\end{figure}

\begin{figure}[htb]
\centering\includegraphics[width=0.9\linewidth]{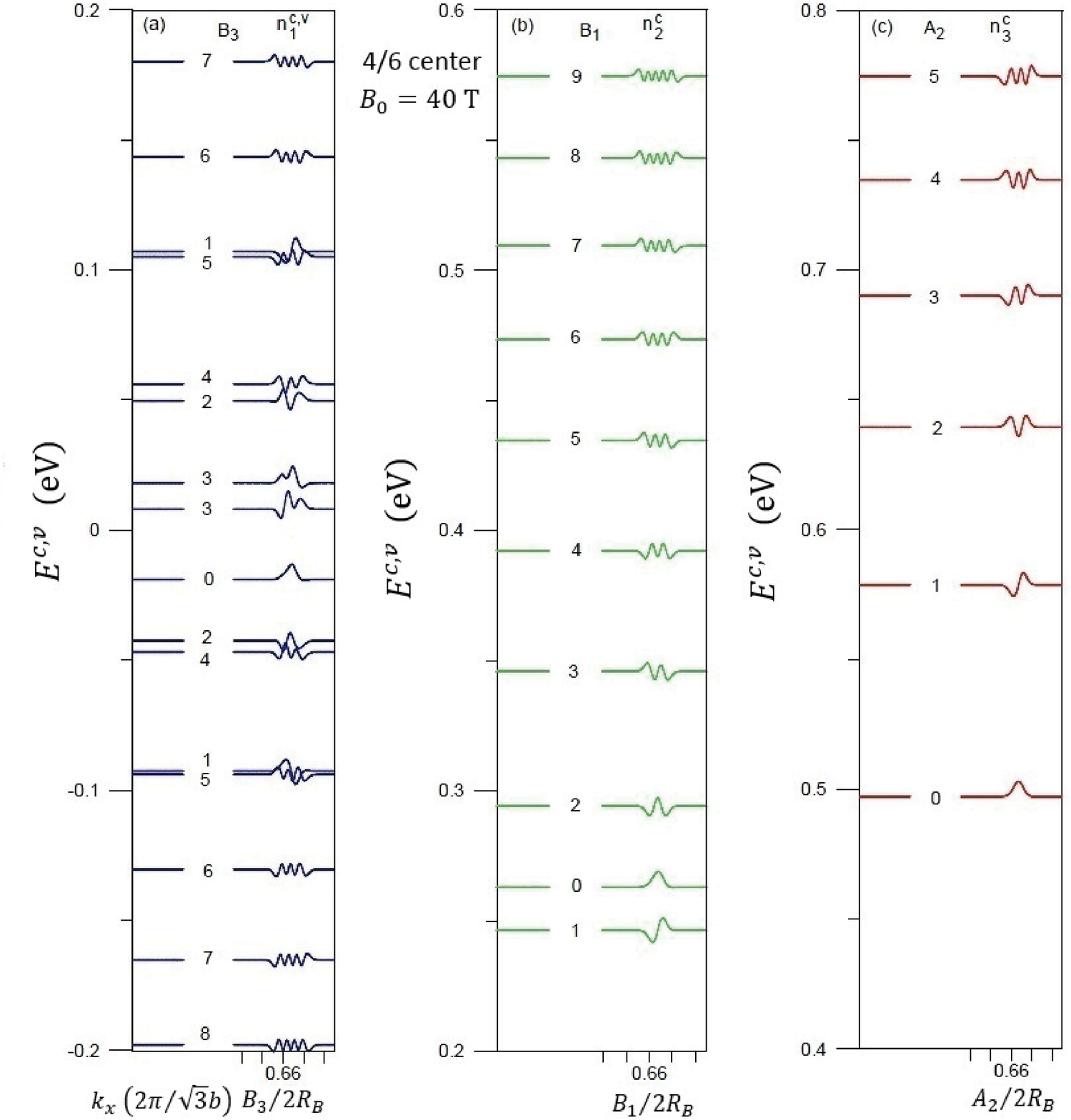}
\caption{The distributions of the LL wavefunctions based on the dominating sublattices centered at the (a)-(c) 4/6 localization under $B_0$ = 40 T.}
\end{figure}

\begin{figure}[htb]
\centering\includegraphics[width=0.9\linewidth]{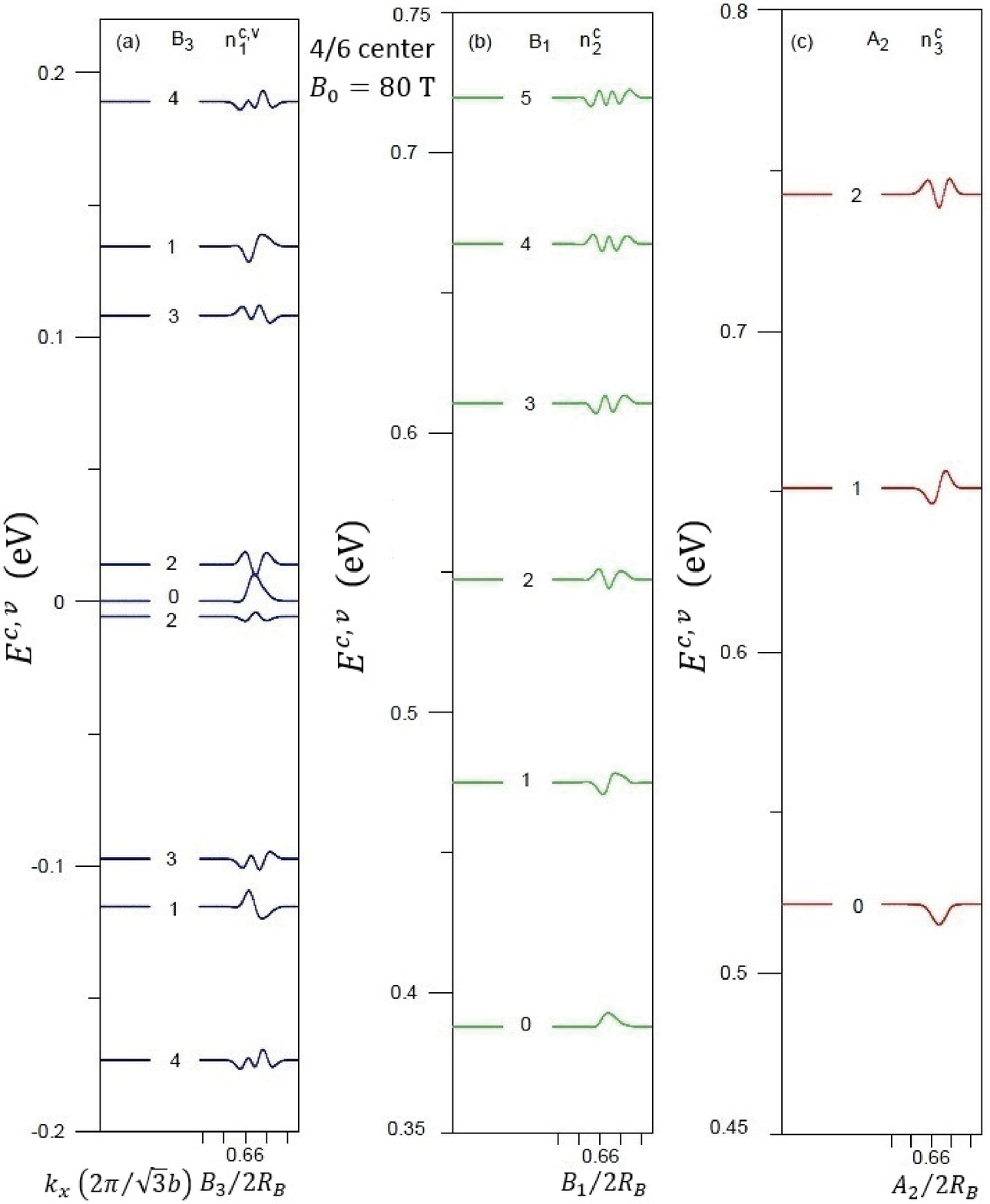}
\caption{The distributions of the LL wavefunctions based on the dominating sublattices centered at the (a)-(c) 4/6 localization under $B_0$ = 80 T.}
\end{figure}

\begin{figure}[htb]
\centering\includegraphics[width=0.9\linewidth]{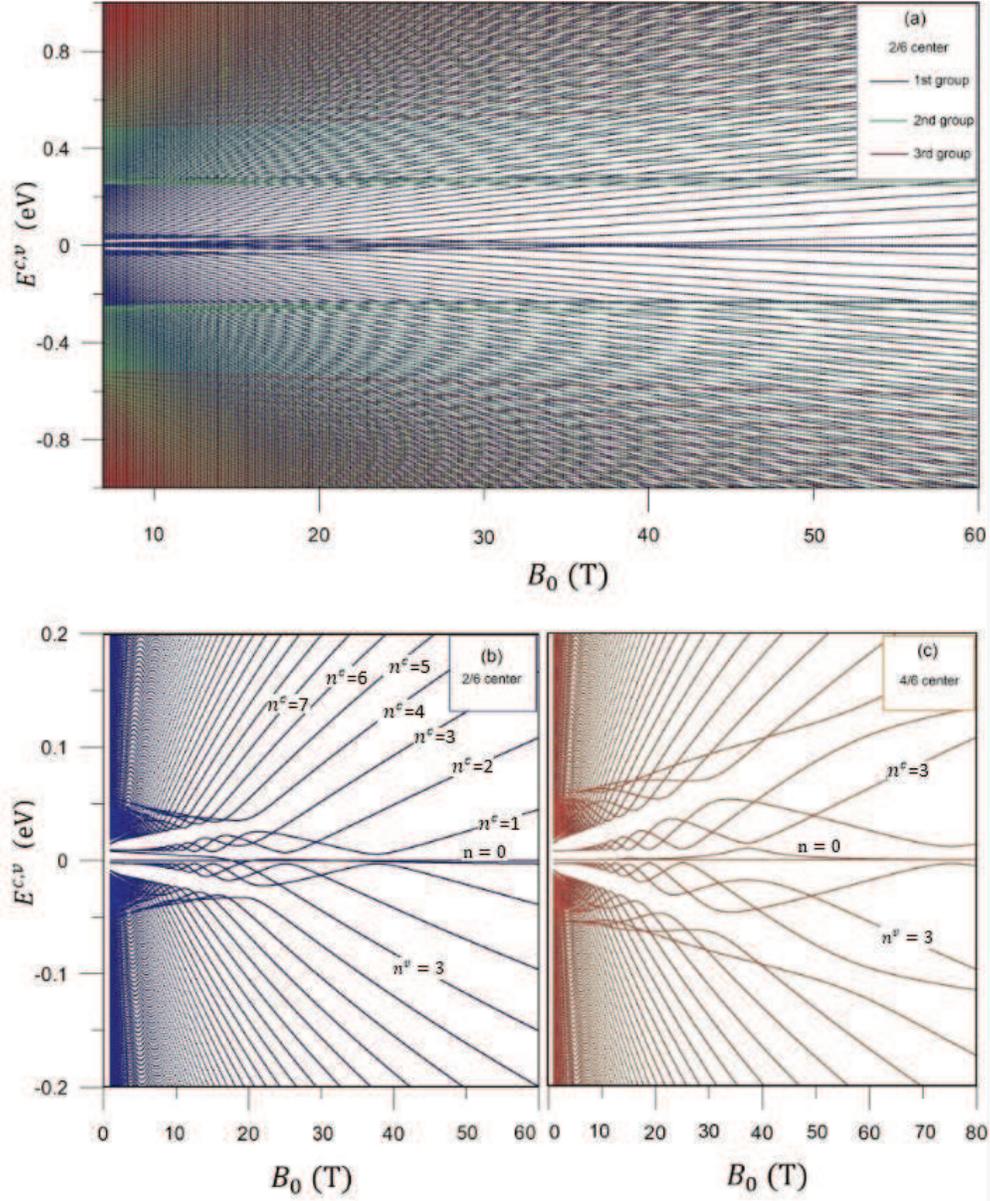}
\caption{Three groups of field-dependent LL energy spectrum for (a) the 2/6 localization center; the low-lying LL spectra for (b) the 2/6 and (c) 4/6 centers.}
\end{figure}

\begin{figure}[htb]
\centering\includegraphics[width=0.9\linewidth]{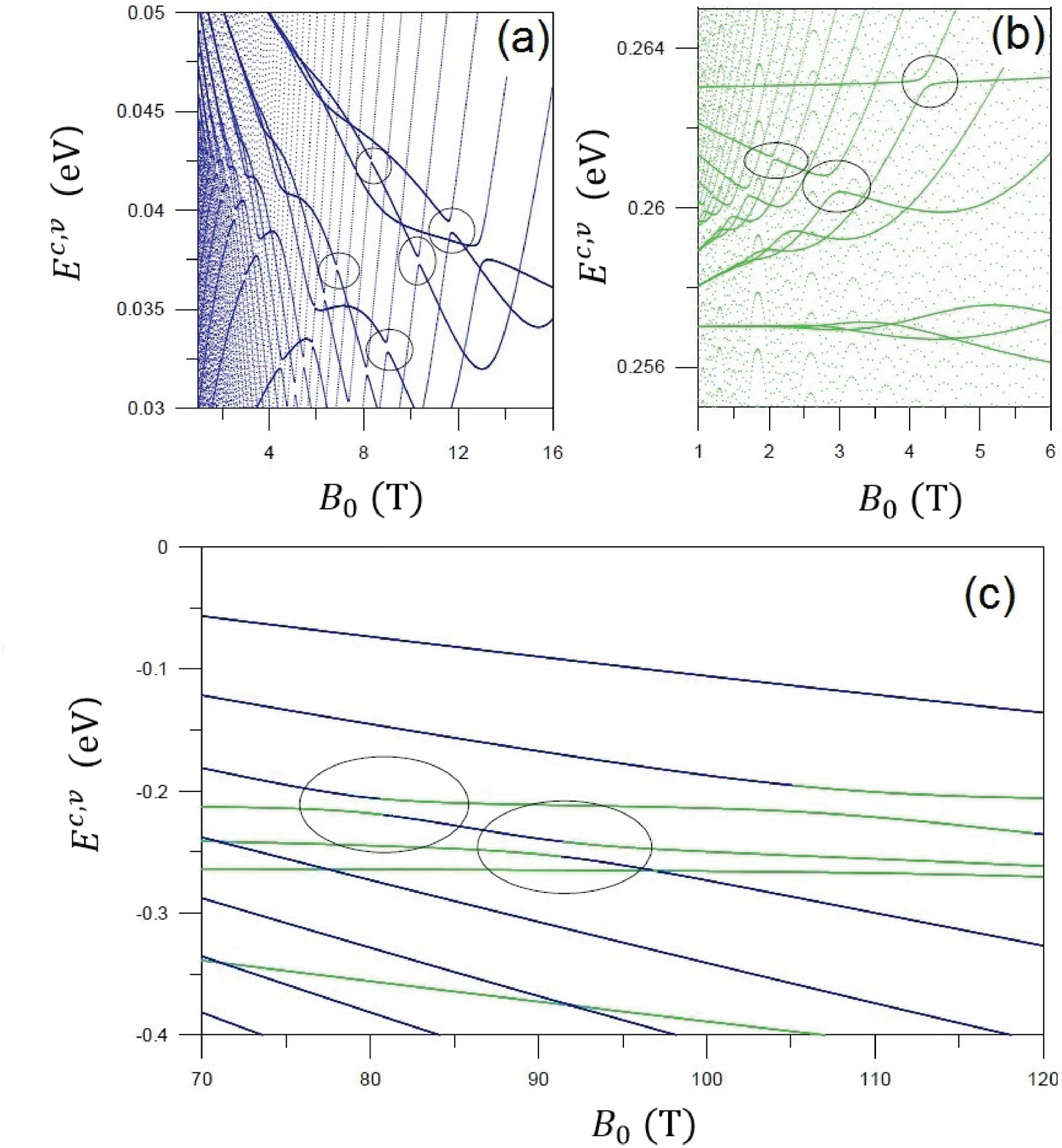}
\caption{The (a) second region of intragroup LL anticrossings in the first group, the (b) frequent intragroup LL anticrossings of  the second group at low field strength, and the (c) intergroup LL anticrossings of the first and second groups.}
\end{figure}

\begin{figure}[htb]
\centering\includegraphics[width=0.9\linewidth]{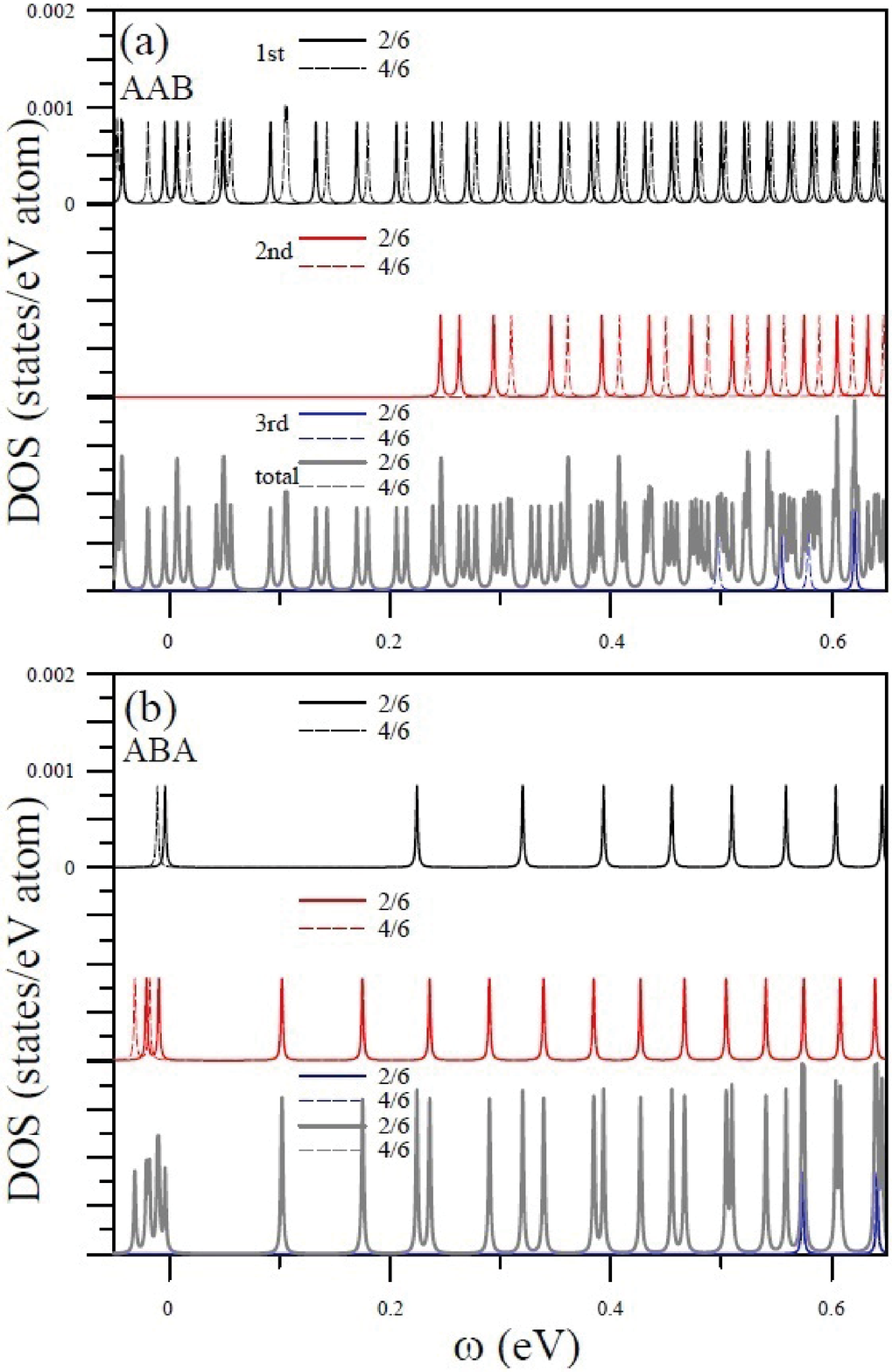}
\caption{The DOS of (a) AAB- and (b) ABA-stacked trilayer graphene at $B_0=40$ T.}
\end{figure}

\begin{figure}[htb]
\centering\includegraphics[width=0.9\linewidth]{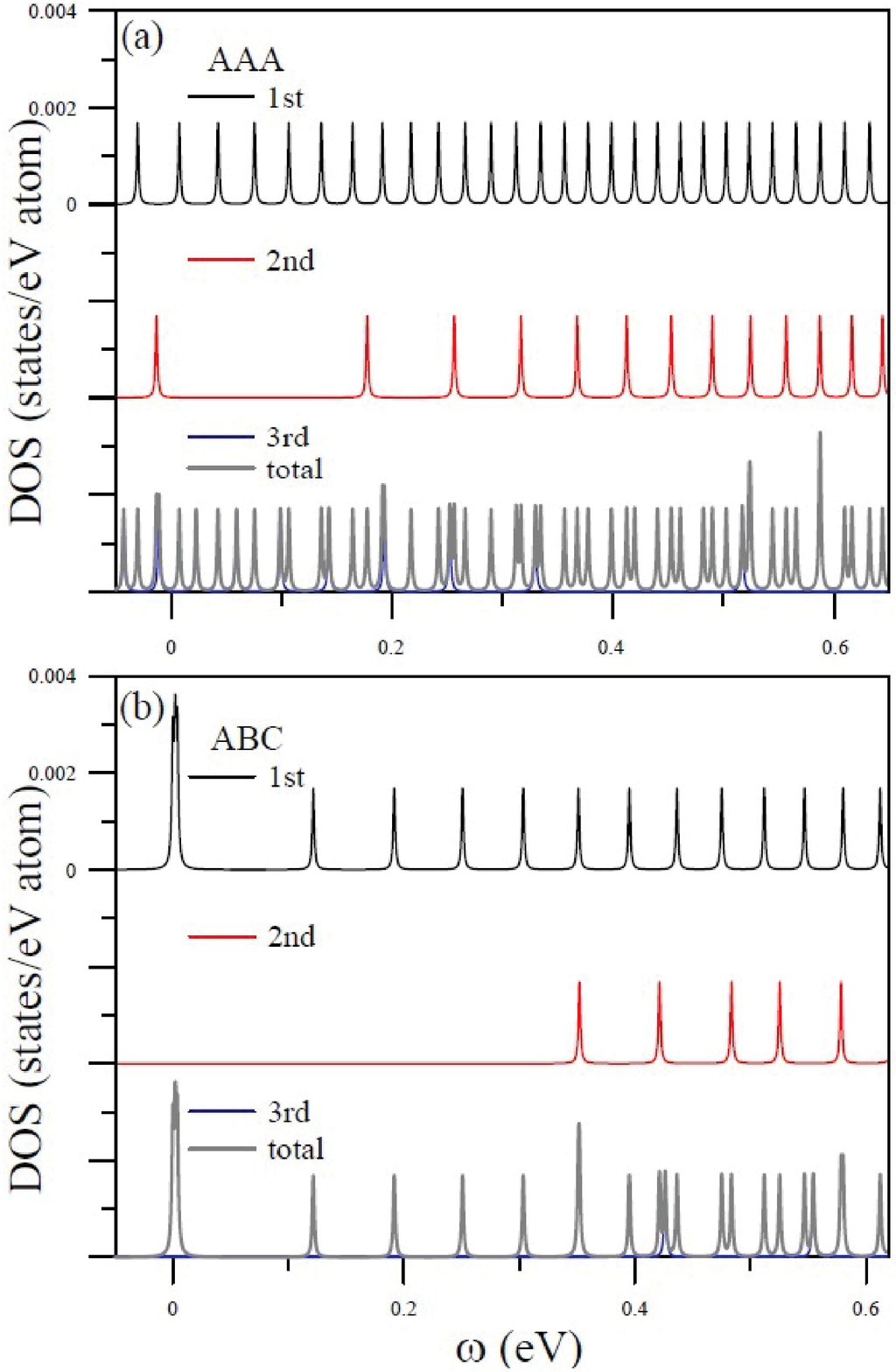}
\caption{The DOS of (a) AAA- and (b) ABC-stacked trilayer graphene at $B_0=40$ T.}
\end{figure}

\end{document}